\documentclass[conference]{IEEEtran}
\IEEEoverridecommandlockouts

\ifCLASSINFOpdf
\else
   \usepackage[dvips]{graphicx}
\fi
\usepackage{url}

\usepackage{graphicx}
\usepackage{algorithm,algpseudocode}

\algnewcommand{\Initialize}[1]{%
\Statex \textbf{Initialize:}
\Statex \hspace*{\algorithmicindent}\parbox[t]{.8\linewidth}{\raggedright #1}
}

\algnewcommand{\Forward}[1]{%
\Statex \textbf{Forward pass:}
\Statex \hspace*{\algorithmicindent}\parbox[t]{.8\linewidth}{\raggedright #1}
}
\usepackage{amsmath,amssymb}
\usepackage{hyperref}
\usepackage{balance}
\usepackage[inline]{enumitem}
\usepackage[super]{nth} % Package for ordinal numbers (1st, 2nd, etc.)

\usepackage{xcolor}

\definecolor{tabblue}{HTML}{1f77b4}
\definecolor{taborange}{HTML}{ff7f0e}
\definecolor{tabgreen}{HTML}{2ca02c}
\definecolor{tabred}{HTML}{d62728}
\definecolor{tabpurple}{HTML}{9267bf}

\begin{document}
%\bstctlcite{IEEEexample:BSTcontrol}
%\abovedisplayskip=3pt plus 3pt
%\belowdisplayskip=3pt plus 3pt
%\abovedisplayshortskip=3pt plus 3pt
%\belowdisplayshortskip=3pt plus 3pt

%\newcommand*{\st@baselinestretch}{0.97}

\title{Channel charting based beamforming}

\author{
	\IEEEauthorblockN{
		Luc Le Magoarou\IEEEauthorrefmark{1}\IEEEauthorrefmark{2}, Taha Yassine\IEEEauthorrefmark{1}\IEEEauthorrefmark{2},  Stéphane Paquelet\IEEEauthorrefmark{1} and Matthieu Crussière\IEEEauthorrefmark{1}\IEEEauthorrefmark{2}
		}
	\IEEEauthorblockA{
		\IEEEauthorrefmark{1}b\raisebox{0.2mm}{\scalebox{0.7}{\textbf{$<>$}}}com, Rennes, France
		}
	\IEEEauthorblockA{
		\IEEEauthorrefmark{2}Univ Rennes, INSA Rennes, CNRS, IETR - UMR 6164, F-35000 Rennes
	}
	}
\maketitle

\begin{abstract}
Channel charting (CC) is an unsupervised learning method allowing to locate users relative to each other without reference. From a broader perspective, it can be viewed as a way to discover a low-dimensional latent space charting the channel manifold. In this paper, this latent modeling vision is leveraged together with a recently proposed location-based beamforming (LBB) method to show that channel charting can be used for mapping channels in space or frequency. Combining CC and LBB yields a neural network resembling an autoencoder. The proposed method is empirically assessed on a channel mapping task whose objective is to predict downlink channels from uplink channels.
\end{abstract}


\IEEEpeerreviewmaketitle

\section{Introduction}
\label{sec:introduction}

\IEEEPARstart{M}{achine} learning techniques have been applied with success to several wireless communication problems in recent years. One can cite for example detection \cite{Samuel2019, He2020}, channel estimation \cite{He2018, Balevi2020, Yassine2022, Chatelier2022} and even end-to-end system learning \cite{Oshea2017, Aoudia2019} (see \cite{Zhang2019, Wang2017} for exhaustive surveys on this topic). However, most of these techniques fall within the supervised learning paradigm and thus require labeled data to operate. The acquisition of such data may be impractical or complex within existing communication systems.

For instance, location-based beamforming (LBB) \cite{Kela2016} is a method allowing to choose precoders based on users' locations, thus suppressing the need for base stations to perform channel estimation or beam sweeping. LBB has recently been tackled successfully with supervised machine learning techniques \cite{Lemagoarou2022}. However, the main drawback of LBB is that it requires having access to the precise users locations, which are difficult to acquire within current systems. 

On the other hand, channel charting (CC) \cite{Studer2018} recently appeared as an \emph{unsupervised} method allowing to locate users relative to each other, based solely on channel knowledge. Predicting this way the relative locations of users from channel measurements has many potential applications, ranging from SNR prediction \cite{Kazemi2020} and pilot reuse \cite{Ribeiro2020}, to  user grouping, proactive handover management or beam-finding (see \cite{Studer2018} for more details on potential applications). CC is an active area of research \cite{Lemagoarou2021} \cite{Ferrand2021} \cite{Rappaport2021} \cite{Yassine2022b} and rapid progress is expected in this direction. The channel chart can actually be seen as a pseudo-location of the users. In this regard, channel charting could be used to relax the constraints of LBB, by replacing the needed location (difficult to estimate) by a pseudo-location obtained by channel charting.

\subsection{Contributions}
In this paper, channel charting is taken as input for location based beamforming. This allows a base station to choose appropriate precoders based on chart locations instead of spatial locations. This alleviate the need for a precise estimation of user locations and opens the way to several applications such as channel mapping in space and frequency \cite{Alrabeiah2019}. The proposed method is empirically assessed on realistic synthetic channels, clearly showing the potential of the approach. 

\subsection{Related work}
To the best of the authors' knowledge, this paper is the first to propose a combination between CC and LBB. The approach relies on previous work done in both areas. The CC method chosen here is inspired by the ones of \cite{Lemagoarou2021,Yassine2022b}. The used LBB method is very close to the one of \cite{Lemagoarou2022}, despite the fact that its input is different (chart locations instead of spatial locations). On the application side, one contribution pursues an objective quite similar to this paper \cite{Ponnada2021}. Indeed, it is proposed in \cite{Ponnada2021} to learn a mapping from chart locations to beam indices. However there are major differences between this approach and the present paper. First of all, the beam is taken from a discrete set of candidate beams in \cite{Ponnada2021} (classification problem) whereas a continuous beamforming is sought here (infinite number of beams, regression problem). Second, features of much higher dimension than the channel vectors and requiring temporal averaging (second order moments) are used in  \cite{Ponnada2021}, whereas an especially designed distance measure that operates directly on instantaneous channel vectors (thus much less complex) is used here. Finally, the neural network used for the mapping is generic (multilayer perceptron) in \cite{Ponnada2021}, whereas a specialized architecture based on random Fourier features is used here.

\begin{figure}[tbp]
\center
\includegraphics[width=0.95\columnwidth]{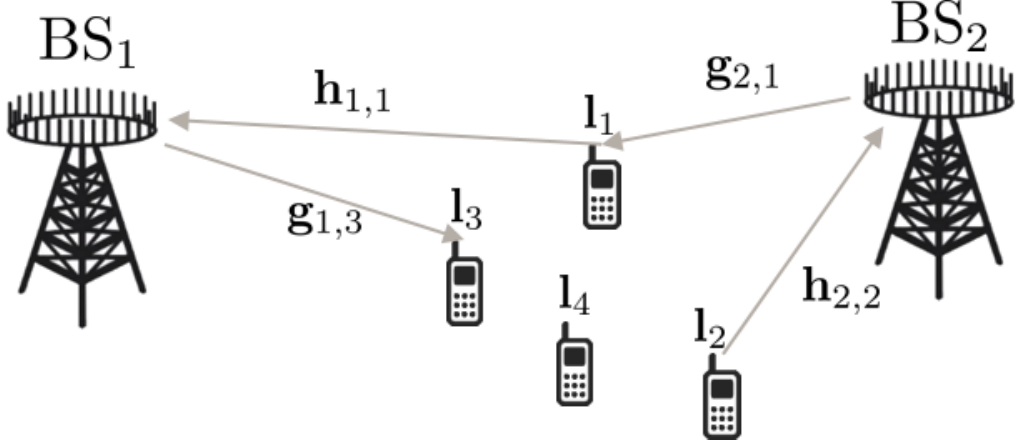}
\caption{Considered setting with $B=2$ and $U=4$.}
\label{fig:setting}
\end{figure}

\section{Problem formulation}
Let us formulate the problem by describing the studied system and stating clearly the pursued objective.
\label{sec:problem}
\subsection{Setting}
\label{sec:setting}
A cell-less multi-user massive MIMO network is considered, in which $B$ base stations (BS) each comprising $A$ antennas communicate over $S$ subcarriers with $U$ user equipments (UE) each having a single antenna. The location of the $j$th user equipment is denoted $\mathbf{l}_j$, and the vectorized uplink (respectively downlink) channel between it and the $i$th base station is denoted $\mathbf{h}_{i,j} \in \mathbb{C}^{AS}$ (respectively $\mathbf{g}_{i,j} \in \mathbb{C}^{AS}$). Note that the uplink and downlink channel are not necessarily over the same band. The considered setting is shown on Fig.\ref{fig:setting}.

\subsection{Objective}
The goal of the proposed contribution is to map channels in space or frequency, which amounts to determine a downlink precoder $\mathbf{w}_{k,j}$ that should be highly correlated with the downlink (target) channel $\mathbf{g}_{k,j}$, based on the knowledge of the uplink (origin) channel $\mathbf{h}_{i,j}$. The origin and target channels are not necessarily on the same band nor even on the same base station ($k$ and $i$ may be different).

\noindent{\bf Performance measure.}
In order to evaluate precoders, the normalized correlation between the precoder $\mathbf{w}_{k,j}$ and the target channel $\mathbf{g}_{k,j}$ is used. It is expressed as
\begin{equation}
\eta_{k,j} \triangleq \frac{|\mathbf{w}_{k,j}^H\mathbf{g}_{k,j}|^2}{\left\Vert \mathbf{g}_{k,j} \right\Vert_2^2}.
\label{eq:corr}
\end{equation}
It is between zero and one (for a perfect precoder), and is tightly linked to the downlink channel capacity (considering a single user), whose expression is 
$$
\log(1+\eta_{k,j}.\text{SNR}_{\text{opt}}),
$$
for received signal of the form $y = \sqrt{P}\mathbf{w}_{k,j}^H\mathbf{g}_{k,j}s+n$ where $n\sim \mathcal{N}(0,\sigma^2)$ is additive noise, $s$ is the sent symbol and $P$ the transmit power. In that setting, $\text{SNR}_{\text{opt}} \triangleq \frac{P\left\Vert \mathbf{g}_{k,j} \right\Vert_2^2}{\sigma^2}$ is the highest achievable signal to noise ratio (SNR)  \cite{Bjornson2017}. In summary, the correlation is a single number between zero and one allowing to determine the maximum achievable downlink spectral efficiency for any transmit power and noise variance considering a given precoder.

\section{Proposed method}
\label{sec:method}
In this section, the inference and training phases of the proposed channel mapping method are described in details. 

\subsection{Inference}

\begin{figure}[tbp]
\center
\includegraphics[width=0.8\columnwidth]{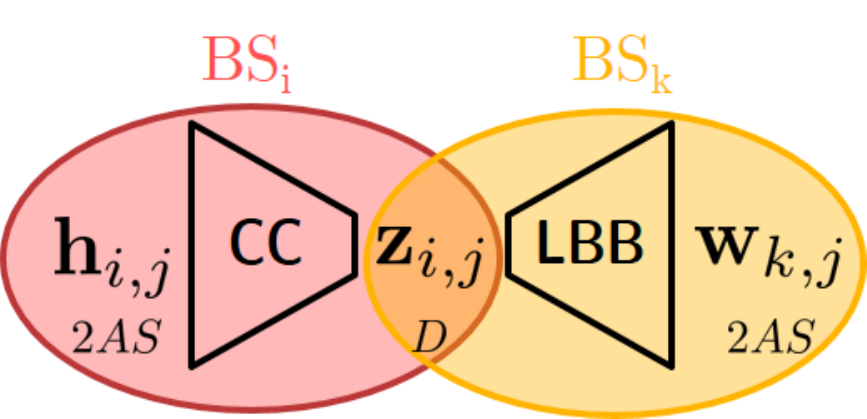}
\caption{Schematic view of the proposed method (the real dimension being shown below each variable).}
\label{fig:method}
\end{figure}

The proposed method comprises three steps:
\begin{enumerate}
\item At first, $D$ latent variables corresponding to the vector $\mathbf{z}_{i,j} \in \mathbb{R}^D$ are computed from the origin channel $\mathbf{h}_{i,j}$ using channel charting methods \cite{Studer2018,Lemagoarou2021} (with $D \ll SA$) at the $i$th BS. In particular, the efficient channel charting method of \cite{Lemagoarou2021} is used in this paper. 
\item Then, the latent variables $\mathbf{z}_{i,j}$, that can be seen as a compressed version of the channel are sent from the $i$th BS to the $k$th BS. 
\item Finally, the $k$th BS performs location based beamforming using the latent variables as inputs, so as to output a precoder $\mathbf{w}_{k,j}$.
\end{enumerate} 
These three steps are depicted on Fig.~\ref{fig:method}. It is important to notice that the dimension is greatly reduced when performing channel charting, which limits the overhead induced by transmitting the latent variables from the $i$th to the $k$th BS.

\subsection{Training} 
\label{ssec:training}

In order to calibrate the CC and LBB steps, a training phase is required.

\begin{figure}[tbp]
\centering
\includegraphics[width=\columnwidth]{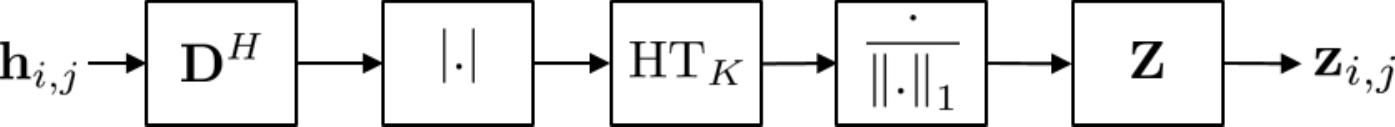}
\caption{Proposed similarity-based neural network for the CC step.}
\label{fig:structure_CC}
\end{figure}

The CC step is calibrated using the method proposed in \cite{Lemagoarou2021}. A phase-insensitive distance between training channels $$\{\mathbf{h}_{i,n}\}_{n=1}^N$$ is first computed and then used within the Isomap nonlinear dimensionality reduction method to obtain chart locations $$\{\mathbf{z}_{i,n}\}_{n=1}^N.$$ The resulting matrices $\mathbf{D} \triangleq (\mathbf{h}_{i,1},\dots,\mathbf{h}_{i,N}) \in \mathbb{C}^{AS\times N}$ and $\mathbf{Z} \triangleq (\mathbf{z}_{i,1},\dots,\mathbf{z}_{i,N}) \in \mathbb{R}^{D\times N}$ are then used to form the feedforward neural network shown on Fig.~\ref{fig:structure_CC} used at inference time to rapidly compute chart location (as first proposed in \cite{Yassine2022b}).

The LBB step is calibrated using the method proposed in \cite{Lemagoarou2022}, except that the inputs are here chart locations instead of true spatial locations. To do so, a database containing chart locations and associated target channels has to be built (classical downlink channel estimation methods have to be used here): 
$$
\{\mathbf{z}_{i,n},\mathbf{g}_{k,n}\}_{n=1}^N.
$$
Then, a neural network comprising a random Fourier features (RFF) layer (see detailed architecture in \cite{Lemagoarou2022}) is trained with a cost function expressed as
\begin{equation}
\mathsf{CF}_k \triangleq 1-\frac{1}{N}\sum_{n=1}^N\frac{|\mathbf{w}_{k,n}^H\mathbf{g}_{k,n}|^2}{\left\Vert \mathbf{g}_{k,n} \right\Vert_2^2}.
\label{eq:cost}
\end{equation}
This cost measures the misalignment of the predicted precoders with respect to the training target channels and is between $0$ (if the precoders are perfectly aligned with the target channels) and $1$ (if the precoders are orthogonal to the target channels).

\begin{figure}[tbp]
\centering
\includegraphics[width=0.55\columnwidth]{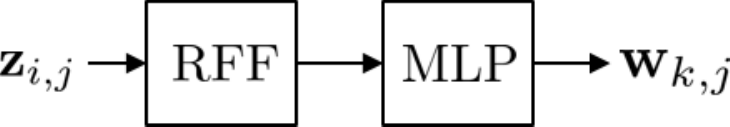}
\caption{Proposed RFF-based neural network for the LBB step.}
\label{fig:network}
\end{figure}
\subsection{Potential applications} The proposed method can be applied to several tasks, depending on the source and target channels. For example, if CC and LBB are done in the same base station ($i=k$) but the frequency bands are different for the uplink (source) and downlink (target) channels, then channel mapping in frequency is carried out. In the more general case where CC and LBB are done in two distinct base stations ($i\neq k$), the proposed method allows to determine a precoder at one BS relying on a channel estimated at another BS (potentially in another frequency band), so that no channel estimation is required at the precoding BS. This is particularly interesting if the CC done at one BS is used to perform LBB at several other BSs (say $B-1$), since in that case channel estimation would be needed at only one BS to determine precoders at $B$ BSs. The overhead due to channel estimation would then be reduced by a factor of $B$. Moreover, the backhaul requirements are kept reasonable, since only $D$ real numbers (dimension of the channel chart which is much smaller than the dimension of the channel) have to be shared for each user.

%\begin{figure}[tbp]
%\centering
%\includegraphics[width=0.5\columnwidth]{network.png}
%\caption{Proposed RFF based neural network for the LBB step (with $Q=4$).}
%\label{fig:network}
%\end{figure}

\section{Experiments}
\label{sec:expe}
To evaluate the proposed approach, a set of experiments is conducted on synthetic data and the results are analysed in this section.

\subsection{Simulation settings}
Multipath channel vectors are generated using the DeepMIMO dataset \cite{Alkhateeb2019}. It consists of channels obtained using a ray-tracing software in an urban outdoor scenario ('O1') (c.f. \cite[Figure 2]{Alkhateeb2019}). Two base stations (i.e. $B=2$) on opposite sides of a street communicate with $U=10000$ UEs spread across the map. The first BS corresponds to BS 16 in the original scenario and the second one corresponds to BS 15. Both BSs are equipped with $A=8\times8$ UPAs. The first one operates at $3.5\,\text{GHz}$ while the second one operates at $28\,\text{GHz}$ and both transmit over sixteen subcarriers (i.e. $S=16$) spanning a bandwidth of $20\,\text{MHz}$. Note that for simplicity, the precoding results reported in this paper are for a single subcarrier (the central one).
All in all, the obtained dataset allows to have access to a triplet $\{\mathbf{h}_{1,j},\mathbf{g}_{2,j},\mathbf{l}_j\}$ for each UE (uplink channel to the first BS, downlink channel from the second BS, user location).

\subsection{Implementation details}
The PyTorch library \cite{Paszke2019} is used to implement and conduct the experiments, and implicitly handles the intricacies of neural network training (gradient computation, backpropagation, etc.). Complex vectors are converted to their real representation by stacking the real and imaginary parts resulting in vectors of twice the original size. The dataset is split into a training set and a test set. The training set size is set at $N=\%70\times U=7000$.

\noindent{\bf CC.} The channel charting step could be performed in one of two manners: in a \textit{one-shot} (to form a baseline) or \textit{on the fly} (as proposed). The one-shot variant is obtained by applying Isomap to the distance matrix of all of the $U$ channels. Only then is the chart split into the training and test sets. On the other hand, the on the fly variant is obtained by first computing matrices $\mathbf{D}$ and $\mathbf{Z}$, restricted to the training set of size $N$, that serve to initialize the NN. The test set is then obtained by computing the chart for the rest of the dataset using the latter. The neighborhood for Isomap is of size 5 for both variants.

The on-the-fly setting has the obvious advantage of being independent from training once performed, but is generally less robust than its one-shot counterpart.

\noindent{\bf LBB.} The input is of size $D=5$ (i.e. dimension of the learned chart), unless otherwise indicated. The number of RFF is set to 600. The MLP consists of 4 layers. Each layer but the last one contains 300 neurons followed by a ReLU non-linearity. The model is trained for 30 epochs using the Adam optimizer \cite{Kingma2014} with a batch size of 100. Finally, the model outputs a  vector $\mathbf{w}\in\mathbb{R}^{2A}$. For simplicity and to limit the complexity of the model in these initial experiments, the task is restricted to the prediction of the precoder corresponding to the central subcarrier only. Extension to the prediction of the full precoder is trivial but likely requires increasing the number of parameters of the NN.

\subsection{Compared variants}

\label{sec:variants}
Five variants of the model are considered for comparison.
\begin{itemize}
    \item LBB is performed at the first BS using the real UEs' locations. Performance is evaluated using the uplink channels. This serves as a \textbf{baseline}.
    \item CC (one-shot) and LBB are both performed at the \nth{1} BS and chained together as described in \ref{sec:method}. The uplink channels are fed as input for training and used to evaluate the performance.
    \item CC (one-shot) is performed first at the \nth{1} BS on the uplink channels. The learned chart is then transmitted to the \nth{2} BS for the LBB step on the downlink channels. This variant corresponds to the channel mapping setting.
    \item This variant is similar to the second one, except this time the learned chart is of dimension $D=3$. The chart is computed in a one-shot.
    \item This variant is similar to the third one, except this time CC is computed on the fly. This last variant corresponds to the most realistic setting that is described in section~\ref{ssec:training}.
\end{itemize}

\subsection{Results}
\begin{figure}[tbp]
\center
\includegraphics[width=1\columnwidth]{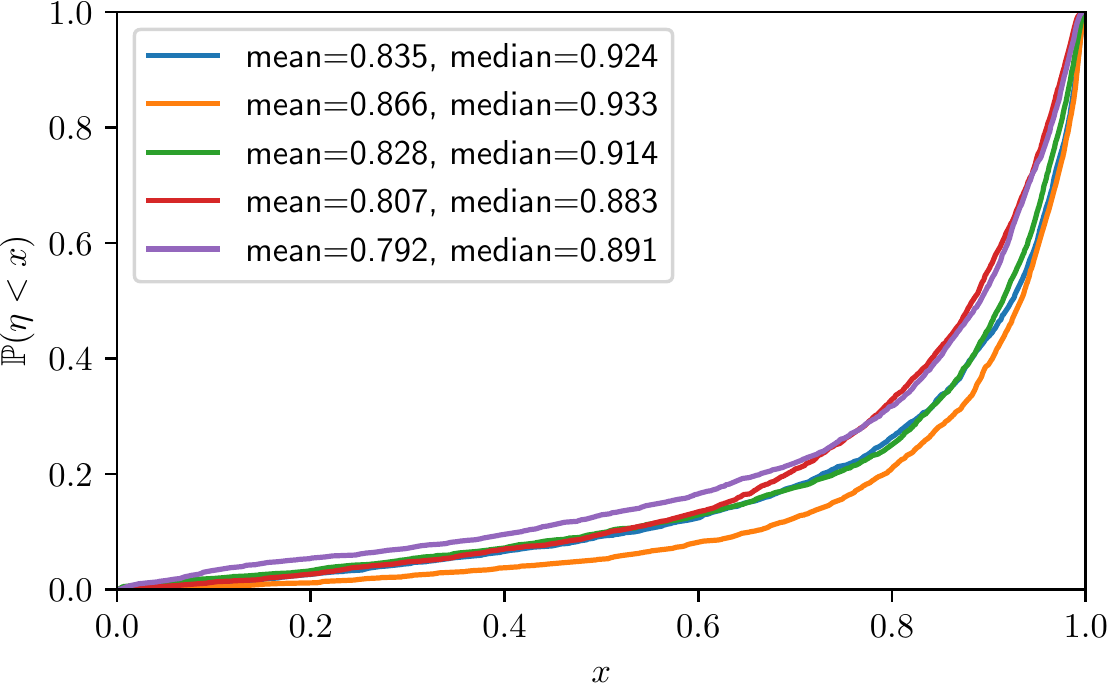}
\caption{CDF of the correlations. \textcolor{tabblue}{Blue} curve corresponds to the original LBB at BS1. \textcolor{taborange}{Orange} curve corresponds to CC (one-shot) and LBB both at BS1 at the same frequency. \textcolor{tabgreen}{Green} curve corresponds to CC (one-shot) at BS1 and LBB at BS2 at different frequencies. \textcolor{tabred}{Red} curve corresponds to CC and LBB both at BS1 at the same frequency with a chart of dimensionality 3. \textcolor{tabpurple}{Purple} curve corresponds to CC (on the fly) at BS1 and LBB at BS2 at different frequencies.}
\label{fig:cdf}
\end{figure}

The five variants of the model are evaluated on the test set. Fig. \ref{fig:cdf} shows the cumulative distribution function (CDF) of the correlations $\eta$---the closer to a Dirac at $x=1$ the better. All methods perform relatively well and achieve a satisfactory level of performance. Interestingly, the CC based LBB variant performs better than the classical one, meaning that the 5 dimensions of the learned chart allow to convey more information than the 3-dimensional spatial locations. This hypothesis is consolidated when comparing classical LBB to the CC based one when keeping only 3 dimensions to the chart where we observe a small decline in performance. Furthermore, the figure shows that in the channel mapping setting, the model performs slightly worse but is on par with the classical one. Finally, the CC based LBB model where the chart is computed on the fly performs the worst, which is expected and highlights the influence of the quality of the learned chart on the subsequent tasks. It worth noting that in this last case, an end-to-end training of the network (i.e., training both the CC NN and the LBB NN) would certainly improve the performance.

\begin{figure}[tbp]
\center
\includegraphics[width=1\columnwidth]{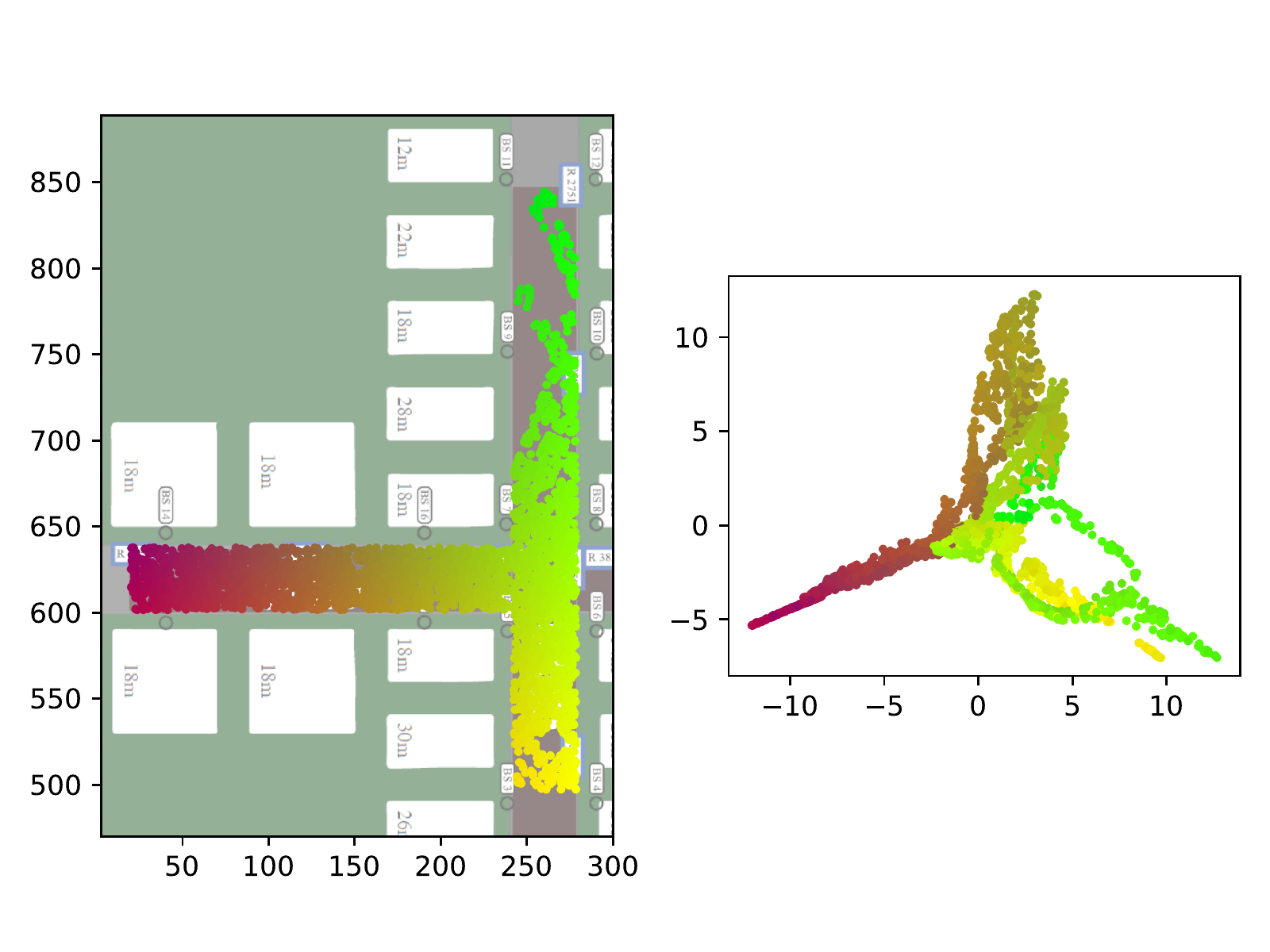}
\caption{Spatial distribution of the UEs in the test set (\textbf{left}) and their corresponding learned chart obtained on the fly (\textbf{right}). Color is added to distinguish local neighborhoods.}
\label{fig:chart}
\end{figure}

\begin{figure}[tbp]
\center
\includegraphics[width=1\columnwidth]{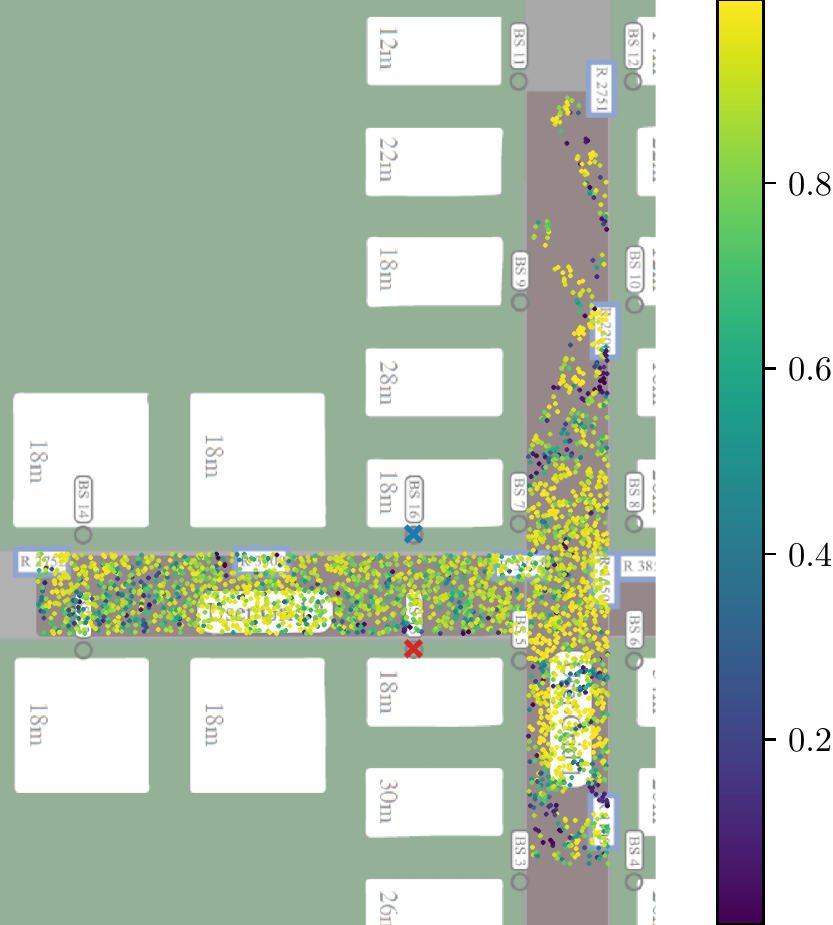}
\caption{Spatial distribution of the correlations. The blue and red crosses correspond to the \nth{1} and \nth{2} BS respectively.}
\label{fig:spatial_perf}
\end{figure}

Fig. \ref{fig:spatial_perf} shows the spatial distribution of the correlations for the last approach described in \ref{sec:variants}. The model succeeds in learning an adequate precoder for most of the spatial locations for both LoS and NLoS channels. Some areas show a high concentration in blue dots (top left and bottom left of the map) corresponding to precoders of low quality and could be explained by locations where propagation conditions with respect to the two base stations are very different (very weak channel between the user and one of the BSs).

\section{Conclusion and outlook}
\label{sec:conclusion}
In this paper, a way to directly predict precoders from channel chart locations was proposed. It amounts to combining wireless channel charting and location-based beamforming. It has the potential of greatly reducing the overhead due to channel estimation in cell-less systems, since CC can be carried out in a base station, and LBB in another one (or several others) that does not need to perform channel estimation nor beam sweeping. The amount of data to transfer between the two BS is reduced since the channel chart is of very low dimension. The proposed method was empirically assessed on realistic synthetic urban channels, showing the great potential of the approach, since it was found that the chart information (in dimension five) was as good as the real location information.

In future work, it would be interesting to train the whole system end-to-end (CC and LBB), which raises the question of how to handle the transmission of gradients and parameters between base stations (decentralized federated learning). Moreover, a theoretical motivation based on the channel manifold in order to set hyperparameters in a principled way would be valuable. Finally, a similar approach could be envisioned for beam prediction within a finite set (as studied in \cite{Ponnada2021}), since the problem can then be seen as a discrete version of the problem studied here.

\section*{Acknowledgement}
This work has been partly funded by the European Commission through the H2020 project Hexa-X (Grant Agreement no. 101015).
\balance

\bibliographystyle{unsrt}
\bibliography{biblio}

% \appendices

% \section{Triplet gradient propagation}
% \section{Handling of complex vectors in neural networks}

\end{document}